\documentclass[preprint2]{aastex}

\newcommand{\vsini}{$v$sin$i$}

\newcommand{\teff}{$T_{eff}$}
\newcommand{\logg}{log~$g$}

\newcommand{\ms}{ms$^{-1}$}
\newcommand{\kms}{kms$^{-1}$}
\newcommand{\msol}{M$_{\odot}$}
\newcommand{\me}{M$_{\rm{\oplus}}$}
\newcommand{\rhk}{log$R_{\rm{HK}}$}

\shorttitle{Two super-Earths Orbiting HD41248}
\shortauthors{Jenkins et al.}

\begin{document}



\title{Two Super-Earths Orbiting the Solar Analogue HD41248 on the edge of a 7:5 Mean Motion Resonance\thanks{Email: jjenkins@das.uchile.cl}}


\author{J.S. Jenkins$^{1,2}$, M. Tuomi$^{2,3}$, R. Brasser$^4$, O. Ivanyuk$^5$, F. Murgas$^{6,7}$}
\affil{$^1$Departamento de Astronomia, Universidad de Chile, Camino el Observatorio 1515, Las Condes, Santiago, Chile, Casilla 36-D\\
$^2$Center for Astrophysics, University of Hertfordshire, College Lane Campus, Hatfield, Hertfordshire, UK, AL10 9AB\\
$^3$University of Turku, Tuorla Observatory, Department of Physics and Astronomy, V\"ais\"al\"antie 20, FI-21500, Piikki\"o, Finland\\
$^4$Institute for Astronomy and Astrophysics, Academia Sinica, Taipei 10617, Taiwan\\
$^5$Main Astronomical Observatory of National Academy of Sciences of Ukraine, 27 Zabolotnoho, Kyiv 127, 03680, Ukraine\\
$^6$Instituto de Astrof\'isica de Canarias, Via Lactea, E38205, La Laguna, Tenerife, Spain\\
$^7$Departamento de Astrof\'{i}sica, Universidad de La Laguna (ULL),  E-38206 La Laguna, Tenerife, Spain}



\begin{abstract}

The number of multi-planet systems known to be orbiting their host stars with orbital periods that place them 
in mean motion resonances is growing.  For the most part, these systems are in first-order resonances 
and dynamical studies have focused their efforts towards understanding the origin and evolution of such 
dynamically resonant commensurabilities. We report here the discovery of two super-Earths that are close to a second-order 
dynamical resonance, orbiting the metal-poor ([Fe/H]=-0.43~dex) and inactive G2V star HD41248.  We analysed 62 HARPS archival radial 
velocities for this star, that until now, had exhibited no evidence for planetary companions.  Using our new Bayesian Doppler signal 
detection algorithm, we find two significant signals in the data, with periods of 18.357~days 
and 25.648~days, indicating they could be part of a 7:5 second-order mean motion resonance. Both semi-amplitudes 
are below 3\ms and the minimum masses of the pair are 12.3 and 8.6~\me, respectively.  Our simulations found 
that apsidal alignment stabilizes the system, and even though libration of the resonant angles was not seen, the system 
is affected by the presence of the resonance and could yet occupy the 7:5 commensurability, which would be 
the first planetary configuration in such a dynamical resonance.  Given the multitude 
of low-mass multiplanet systems that will be discovered in the coming years, we expect more of these second-order resonant 
configurations will emerge from the data, highlighting the need for a better understanding of the dynamical interactions 
between forming planetesimals.  

\end{abstract}





\section{Introduction}

Resonances seem to be a common feature in nature when bodies with measurable gravitational fields interact dynamically with one and other.  
In the Solar System, various bodies are found to be in mean motion resonances (MMRs), yet even though such objects can be found to have period ratios that are close 
to known MMRs, confirmation of the existence of any MMR can only be made by studying the system dynamically to confirm effects 
such as libration of the resonant angles.  Examples include the 2:5 MMR of Jupiter and Saturn 
(\citealp{michtchenko01}) and the 3:2 MMR between Neptune and Pluto (\citealp{peale76}).  In fact a 1:2 MMR between Jupiter and Saturn may 
have been the driving force behind the current configuration of the solar systems outer planets (\citealp{gomes05}; \citealp{tsiganis05}; 
\citealp{morbidelli05}).

Beyond planetary bodies there are also various moons and asteroids that exhibit MMRs.  A famous example of these comes from Jupiters moons 
Ganymede, Europa, and Io that form a 1:2:4 resonant set (\citealp{peale76}).  Such resonances provide important constraints on the formation and evolution 
of migrating bodies and allow a window into the dynamics of evolving systems.  Therefore, discovering new resonant exoplanetary 
systems can provide unique constraints on the early evolution of exoplanets, along with a contextual view of the whole ensemble of planet formation 
and evolution in general.

Currently we have confirmed a number of exoplanetary systems that have pairs or more of planets in some resonant configuration by radial velocities.  
2:1 resonances seem to be the most common by-product of giant planet formation (e.g. \citealp{marcy01}).  Further resonances have been witnessed, 
including a Jupiter-moon like 1:2:4 orbital resonant set (\citealp{rivera10b}), however these systems tend to be first-order resonances.

\citet{desort08} reported the discovery of the HD60532 exoplanetary system that contains a pair of planets in a second-order MMR.  Again these planets 
were found to be gas giants with masses above a Jupiter-mass and \citet{laskar09} confirmed they are indeed in a 3:1 MMR configuration.  Recently, \citet{fabrycky12} 
used Kepler transit timing measurements to detect resonant systems including that of Kepler-29 that seems to have two planets orbiting the star locked in a 9:7 MMR.  
This system does not have radial velocity information to confirm the nature of the objects, however this could be the first super-Earth planetary system in such 
a dynamical commensurability.  Kepler data also indicate that 3:1 MMRs are the most common second-order resonances (\citealp{lissauer11}).

In this work we present the discovery of a possible new second-order MMR configuration that has never been previously observed for planets.  
We show that the magnetically inactive G2V star HD41248 hosts at least two rocky planets that are on the edge of a 7:5 MMR, by analysing extensive archival data from HARPS 
that previously did not contain any known Doppler signals.  In $\S$~\ref{star} we present the data we have used in this study, then in $\S$~\ref{sun} we discuss the 
comparison between this star and the Sun.  In $\S$~\ref{rv} we present the Keplerian analysis of our radial velocities, and in $\S$~\ref{activity} we show that 
the signals are not associated with any activity related phenomena.  Finally, $\S$~\ref{stability} presents our dynamical stability tests for this system and we finish with a 
discussion of the system in $\S$~\ref{discussion}.

\section{HD41248 Data\label{star}}

All data in this paper were taken from the ESO HARPS Archive\footnote{http://www.eso.org/Archive}, a community tool that allows 
users to download fully reduced and analysed data that has been processed using the HARPS-DRS Version 3.5.  The 
pipeline performs the usual reduction steps for high resolution echelle spectra, from bias and flatfielding, to extraction and 
wavelength calibration.  

A total of 62 velocities for the star HD41248 were downloaded and used in this analysis as part of our project to discover new rocky planets 
orbiting nearby Sun-like stars using our novel methodologies.  The baseline of observations is close to 7.5~years (BJD 2452943.85284 to 
2455647.57967) and in general a high level of data quality was maintained.  The median S/N for the set is $\sim$100 at a wavelength 
of around 6050\AA, with a lowest value of 26 and a highest value of 150.  None of the data we downloaded were rejected from our final analysis.

After reduction and extraction of the observation has finished, post-reduction analysis is also performed on the spectra and this 
consists of cross-correlating each of the echelle orders with a pre-fabricated binary mask to generate an order-per-order 
cross-correlation function (CCF), which are then combined using a weighting scheme to produce 
a single stable mean CCF.  This mean CCF is then fit by a gaussian function and the gaussian model allows the software to generate 
a precise and absolute radial velocity measurement.

\begin{deluxetable}{ccccccccc}
\center
\tabletypesize{\scriptsize}
\tablecaption{HARPS radial velocities for HD41248.}
\label{rv_table}
\tablehead{
\colhead{BJD} & \colhead{DRS-RV} & \colhead{DRS-$\sigma_{\rm{RV}}$} & \colhead{TERRA-RV} & \colhead{TERRA-$\sigma_{\rm{RV}}$} &
\colhead{$S_{\rm{MW}}$} & \colhead{\rhk} & \colhead{BIS}  & \colhead{FWHM}  \\
\colhead{[days]} & \colhead{[\ms]} & \colhead{[\ms]} & \colhead{[\ms]} & \colhead{[\ms]} &
\colhead{[dex]} & \colhead{[dex]} & \colhead{[\ms]}  & \colhead{[\ms]}  \\
}
\startdata
 2452943.8528426 &  3526.591   &     2.588 &  -3.512  &   2.591 &        0.1686$\pm$0.0024&   -4.9348     &      35.92$\pm$3.66     &       6721.78   \\
 2452989.7102293 &  3519.139   &     4.063 &  -7.736  &   4.223    &     0.1566$\pm$0.0045&   -5.0089     &      27.39$\pm$5.74     &       6719.01    \\
 2452998.6898180 &  3526.427   &     5.425 &   1.179   &   6.226 &        0.1495$\pm$0.0062 &   -5.0599     &      33.53$\pm$7.67     &       6701.20     \\
 2453007.6786518 &  3526.626   &     2.525 &  -3.456  &   2.232  &       0.1609$\pm$0.0025&   -4.9810     &      28.61$\pm$3.57     &       6718.19    \\
 2453787.6079555 &  3522.442   &     2.758 &  -3.192  &   3.353  &       0.1566$\pm$0.0031&   -5.0091     &      31.30$\pm$3.90     &       6718.53    \\
 2454055.8375443 &  3523.175   &     2.062 &  -6.636  &   2.226 &        0.1674$\pm$0.0018&   -4.9418     &      23.94$\pm$2.91     &       6714.51    \\
 2454789.7207967 &  3522.987   &     0.817 &  -4.361  &   0.897 &        0.1705$\pm$0.0005&   -4.9241     &      27.43$\pm$1.15     &       6722.19    \\
 2454790.6943362 &  3519.487   &     0.899 &  -6.817  &   0.916  &       0.1698$\pm$0.0005&   -4.9281     &      30.82$\pm$1.27     &       6724.20    \\
 2454791.7055725 &  3522.466   &     0.834 &  -4.765  &   0.831 &        0.1708$\pm$0.0005&   -4.9225     &      29.54$\pm$1.18     &       6720.59    \\
 2454792.7042506 &  3522.290   &     0.795 &  -4.763  &   0.779 &        0.1723$\pm$0.0005&   -4.9143     &      28.09$\pm$1.12     &       6728.64    \\
 2454793.7211230 &  3524.992   &     0.890 &  -2.854  &   0.984 &        0.1727$\pm$0.0005&   -4.9124     &      25.27$\pm$1.25     &       6727.73    \\
 2454794.6946036 &  3527.038   &     0.893 &   0.860   &   1.051 &        0.1715$\pm$0.0005 &   -4.9191     &      29.58$\pm$1.26     &       6732.04    \\
 2454795.7156306 &  3528.445   &     0.913 &   0.874   &   0.909 &        0.1737$\pm$0.0005 &   -4.9070     &      29.54$\pm$1.29     &       6725.91    \\
 2454796.7195391 &  3528.207   &     0.957 &   0.513   &   0.975 &        0.1736$\pm$0.0006 &   -4.9076     &      30.32$\pm$1.35     &       6727.87    \\
 2454797.7051254 &  3528.994   &     0.908 &   3.036   &   1.137  &       0.1750$\pm$0.0005 &   -4.9001     &      27.76$\pm$1.28     &       6733.21    \\
 2454798.6972277 &  3531.198   &     0.915 &   3.831   &   0.906 &        0.1731$\pm$0.0005 &   -4.9101     &      25.13$\pm$1.29     &       6731.22    \\
 2454902.5907553 &  3525.084   &     2.021 &   0.000   &   2.525  &       0.1796$\pm$0.0018 &   -4.8768     &      21.34$\pm$2.85     &       6729.22    \\
 2454903.5172666 &  3527.587   &     0.783 &   0.359   &   0.909 &        0.1722$\pm$0.0004 &   -4.9151     &      27.68$\pm$1.10     &       6726.66    \\
 2454904.5185682 &  3525.760   &     0.901 &  -2.032  &   0.862 &        0.1737$\pm$0.0005&   -4.9071     &      27.18$\pm$1.27     &       6722.67    \\
 2454905.5355291 &  3527.552   &     0.898 &  -0.148  &   0.897 &        0.1706$\pm$0.0005&   -4.9240     &      27.97$\pm$1.27     &       6722.94    \\
 2454906.5179999 &  3527.947   &     1.041 &   0.122   &   1.114 &        0.1713$\pm$0.0006 &   -4.9201     &      31.02$\pm$1.47     &       6723.85    \\
 2454907.5647983 &  3527.517   &     0.905 &   0.140   &   1.123 &        0.1731$\pm$0.0006 &   -4.9101     &      28.88$\pm$1.28     &       6727.04    \\
 2454908.5603822 &  3526.226   &     0.804 &  -2.046  &   0.961 &        0.1721$\pm$0.0004&   -4.9155     &      27.94$\pm$1.13     &      6720.84    \\
 2454909.5380036 &  3527.107   &     0.854 &  -0.513  &   0.928  &       0.1699$\pm$0.0005&   -4.9277     &      25.45$\pm$1.20     &      6721.35    \\
 2454910.5385064 &  3528.236   &     1.115 &   1.119   &   1.170 &        0.1727$\pm$0.0008 &   -4.9121     &      27.50$\pm$1.57     &       6726.82    \\
 2454911.5427244 &  3524.975   &     0.747 &  -3.570  &   0.722 &        0.1715$\pm$0.0004&   -4.9189     &      28.13$\pm$1.05     &       6721.04    \\
 2454912.5392921 &  3525.216   &     0.739 &  -3.038  &   0.898 &        0.1719$\pm$0.0004&   -4.9167     &      25.78$\pm$1.04     &       6718.78    \\
 2455284.5272133 &  3528.228   &     0.734 &  -1.092  &   0.917 &        0.1751$\pm$0.0004&   -4.8998     &      28.42$\pm$1.03     &       6724.46    \\
 2455287.5109103 &  3524.603   &     0.881 &  -4.445  &   0.921 &        0.1743$\pm$0.0005&   -4.9038     &      25.18$\pm$1.24     &       6733.01    \\
 2455288.5285775 &  3523.316   &     0.754 &  -4.407  &   0.807 &        0.1731$\pm$0.0004&   -4.9104     &      25.43$\pm$1.06     &       6730.40    \\
 2455289.5460248 &  3526.698   &     0.789 &  -1.501  &   0.923 &        0.1743$\pm$0.0004&   -4.9036     &      26.29$\pm$1.11     &       6723.59    \\
 2455290.5095380 &  3525.900   &     0.891 &  -0.647  &   0.996 &        0.1737$\pm$0.0005&   -4.9068     &      21.95$\pm$1.26     &       6727.32    \\
 2455291.5216615 &  3526.083   &     1.006 &  -3.054  &   1.060 &        0.1707$\pm$0.0006&   -4.9233     &      26.85$\pm$1.42     &       6734.67    \\
 2455293.5043818 &  3527.359   &     0.974 &   1.254   &   1.113 &        0.1719$\pm$0.0006 &   -4.9165     &      27.39$\pm$1.37     &       6727.22    \\
 2455304.5180173 &  3522.412   &     1.689 &  -0.532  &   5.028  &       0.1318$\pm$0.0038&   -5.2203     &      31.20$\pm$2.38     &      6801.22    \\
 2455328.4550220 &  3529.109   &     0.791 &   2.304   &   0.828 &        0.1746$\pm$0.0005 &   -4.9022     &      20.73 $\pm$1.11     &       6735.38    \\
 2455334.4564390 &  3532.003   &     1.365 &   7.425   &   1.432 &        0.1685$\pm$0.0014 &   -4.9353     &      25.33 $\pm$1.93     &       6737.32    \\
 2455387.9305071 &  3530.983   &     1.039 &   4.143   &   0.986 &        0.1708$\pm$0.0008 &   -4.9225     &      27.29 $\pm$1.47     &       6737.65    \\
 2455390.9312188 &  3531.330   &     1.577 &   4.830   &   1.517 &        0.1635$\pm$0.0013 &   -4.9652     &      30.24 $\pm$2.23     &       6734.97    \\
 2455434.8790630 &  3516.749   &     3.334 &  -4.089  &   3.355 &        0.1479$\pm$0.0036&   -5.0724     &      31.59$\pm$4.71     &       6740.53    \\
 2455439.8843407 &  3527.755   &     3.258 &  10.200  &   4.319 &        0.1488$\pm$0.0038&   -5.0649     &      32.80$\pm$4.60     &       6723.47    \\
 2455445.9241644 &  3522.932   &     3.113 &  -0.217  &   3.839 &        0.1545$\pm$0.0036&   -5.0233     &      28.30$\pm$4.40     &       6727.79    \\
 2455465.8566225 &  3526.615   &     1.412 &   1.415   &   1.440  &       0.1706$\pm$0.0012 &   -4.9237     &      23.18 $\pm$1.99     &       6731.85    \\
 2455480.8795598 &  3527.592   &     1.133 &   1.534   &   1.245 &        0.1706$\pm$0.0008 &   -4.9237     &      27.09 $\pm$1.60     &       6728.47    \\
 2455483.8136024 &  3525.010   &     1.740 &  -1.948  &   1.837 &        0.1704$\pm$0.0015&   -4.9250     &      27.97$\pm$2.46     &       6734.01    \\
 2455488.8262312 &  3525.839   &     0.747 &  -0.835  &   0.818 &        0.1708$\pm$0.0004&   -4.9226     &      31.07$\pm$1.05     &       6725.47    \\
 2455494.8532009 &  3529.808   &     0.954 &   3.114   &   1.123 &        0.1686$\pm$0.0006 &   -4.9348     &      26.88 $\pm$1.34     &       6730.94    \\
 2455513.7823100 &  3528.860   &     1.286 &  -0.462  &   1.438  &       0.1715$\pm$0.0009&   -4.9186     &      35.60$\pm$1.81     &       6740.83    \\
 2455516.7515789 &  3530.744   &     0.936 &   4.112   &   1.032 &        0.1749$\pm$0.0007 &   -4.9005     &      22.71 $\pm$1.32     &       6736.78    \\
 2455519.7046533 &  3528.854   &     1.198 &   3.453   &   1.116 &        0.1682$\pm$0.0009 &   -4.9375     &      21.95 $\pm$1.69     &       6735.81    \\
 2455537.7997291 &  3527.714   &     0.718 &   0.897   &   0.859 &        0.1756$\pm$0.0005 &   -4.8968     &      27.01 $\pm$1.01     &       6730.29    \\
 2455545.7213656 &  3529.198   &     0.854 &   3.099   &   0.990 &        0.1749$\pm$0.0005 &   -4.9006     &      26.68 $\pm$1.20     &       6737.59    \\
 2455549.7548559 &  3527.949   &     0.833 &  -0.567  &   0.896 &        0.1714$\pm$0.0005&   -4.9195     &      31.97$\pm$1.17     &       6734.54    \\
 2455576.7923481 &  3525.346   &     1.014 &  -1.481  &   1.120 &        0.1654$\pm$0.0007&   -4.9533     &      29.17$\pm$1.43     &       6733.68    \\
 2455580.7312518 &  3519.411   &     0.895 &  -7.373  &   0.965 &        0.1688$\pm$0.0006&   -4.9339     &      32.94$\pm$1.26     &       6729.42    \\
 2455589.7734088 &  3528.423   &     1.420 &   0.958   &   1.579 &        0.1684$\pm$0.0013 &   -4.9360     &      24.59 $\pm$2.00     &       6733.73    \\
 2455612.6068850 &  3527.663   &     0.816 &   0.976   &   0.713 &        0.1714$\pm$0.0005 &   -4.9193     &      25.58 $\pm$1.15     &       6731.09    \\
 2455623.6361828 &  3528.301   &     1.175 &   0.455   &   1.316 &        0.1685$\pm$0.0011 &   -4.9357     &      28.96 $\pm$1.66     &       6726.23    \\
 2455629.5528393 &  3528.652   &     0.944 &   1.864   &   1.002 &        0.1682$\pm$0.0006 &   -4.9370     &      25.71 $\pm$1.33     &       6732.34    \\
 2455641.5542106 &  3528.979   &     1.005 &   1.879   &   1.012 &        0.1697$\pm$0.0007 &   -4.9286     &      28.91 $\pm$1.42     &       6727.20    \\
 2455644.5845033 &  3529.351   &     1.260 &   3.245   &   1.381 &        0.1742$\pm$0.0006 &   -4.9043     &      19.09 $\pm$1.78     &       6740.10    \\
 2455647.5796694 &  3529.945   &     1.372 &   1.898   &   1.259 &        0.1669$\pm$0.0012 &   -4.9448     &      29.10 $\pm$1.94     &       6737.56    \\
\enddata
\end{deluxetable}

The values extracted from the ESO Archive DRS processing are also backed up with the velocities generated using the HARPS-TERRA software 
(\citealp{anglada-escude12a}; \citealp{anglada-escude12b}), which we use as a sanity check for stars like the Sun as the DRS values tend to 
produce higher precision for these stars, in comparison to cooler M dwarfs where TERRA works better.  The final Barycentric Julian Dates, DRS, 
and TERRA radial velocities, along with DRS and TERRA uncertainties, are shown in Table~\ref{rv_table}.

\section{HD41248 vs the Sun\label{sun}}

The properties of HD41248 are summarised in Table~\ref{tab:values} but some of the most interesting features are highlighted here.  First of all, 
the star has a spectral type of G2V and $B-V$ colour of 0.624, meaning it is a solar analogue since the solar values are G2V and 0.642, respectively.  
Rocky planets around such stars can provide direct tests of planet formation mechanisms and architectures of systems orbiting Sun-like stars.

\begin{figure}
\vspace{5.5cm}
\hspace{-4.0cm}
\includegraphics{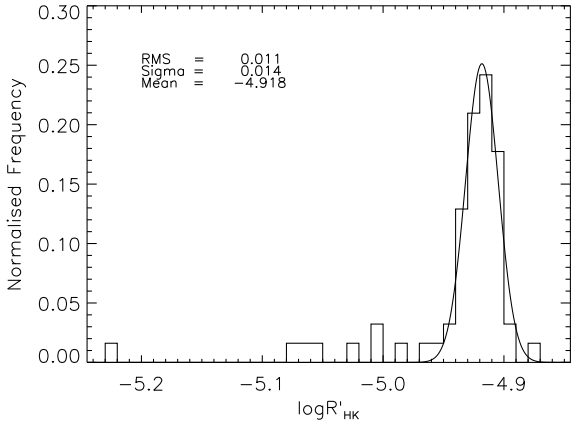}
\vspace{-0.2cm}
\caption{Histogram of the \rhk ~activity indices of the timeseries data for HD41248.  The best fit Gaussian model to the 
activities is represented by the solid curve and the associated data for this model is shown in the plot.}
\label{act_hist}
\end{figure}

\subsection{Chromospheric Activity}

Low amplitude signals are most easily sought after in radial velocity data of the most quiescent and slowly rotating stars.  HD41248 has a 
\rhk ~activity index of -4.92~dex and a rotational velocity of only 2.4$\pm$~\kms, highlighting that this star is an ideal candidate for such studies.  
The Sun has a \rhk ~activity index of -4.91~dex and \vsini ~of 1.6$\pm$0.3~\kms (\citealp{pavlenko12}) showing both these stars have good agreement 
in their evolutionary properties too.

Fig.~\ref{act_hist} shows the distribution of these activity indices and the best fit model Gaussian to the data.  The \rhk ~values were computed using 
the HARPS 1D spectra following the procedures explained in \citet{jenkins06,jenkins08,jenkins11}.  Clearly most of the data are tightly clustered around 
the mean \rhk ~of -4.92~dex.  The distribution is tightly packed, with a scatter of only 0.01~dex, which is both intrinsic variability of the star and 
uncertainty in the measurements.  There are a few values with activities less than -5~dex, however these tend to be the lowest S/N spectra and therefore 
there is a bias in the line core measurements as random noise causes negative read values that artificially draw the line core values lower than they really are, 
and therefore the overall activity index is found to be lower than it should be.

\begin{table}
\center
\scriptsize
\caption{Stellar parameters for HD41248.}
\label{tab:values}
\begin{tabular}{lll}
\hline
\multicolumn{1}{l}{Parameter}& \multicolumn{1}{l}{HD41248} & \multicolumn{1}{l}{Reference} \\ \hline

RA J2000   (h:m:s)                  & 06$^{\rm{h}}$00$^{\rm{m}}$32$^{\rm{s}}$.781 & \citet{perryman97} \\
Dec J2000 (d:m:s)                  &  -56$^{\rm{o}}$09$'$42.61$''$ & \citet{perryman97} \\
Spectral Type                          & G2V  &  \citet{perryman97} \\
$B-V$                                     & 0.624 & \citet{perryman97} \\
$V$                                         & 8.82  & \citet{perryman97} \\
distance (pc)                            & 52.38$\pm$1.95 & \citet{vanleeuwen07} \\
M$_{V}$                                    & 5.22$\pm$0.08 & This Work \\
log$R'$$_{\rm{HK}}$                   & -4.94 & This Work \\
Hipparcos $N$$_{\rm{obs}}$             & 106  & \citet{perryman97} \\
Hipparcos $\sigma$                     & 0.015  & \citet{perryman97}  \\
$\Delta$$M_{V}$                        & -0.700  & \citet{jenkins11} \\
$L_{\rm{\star}}$/$L_{\odot}$           & 0.68$\pm$0.03  & This Work \\
$M_{\rm{\star}}$/$M_{\odot}$     & 0.92$\pm$0.05  & This Work \\
$R_{\rm{\star}}$/$R_{\odot}$           & 0.78$\pm$0.04  & This Work \\
$T$$_{\rm{eff}}$ (K)                     & 5713$\pm$50   & \citet{ivanyuk13} \\
$[$Fe/H$]$                              & -0.43$\pm$0.10  & \citet{ivanyuk13} \\
\logg                                        & 4.48$\pm$0.10  & \citet{ivanyuk13}  \\
U,V,W (km/s)                            & -12.97, -17.81, 26.74 & \citet{jenkins11} \\
$P_{\rm{rot},v~\rm{sin}~i}$ (days)     & 16  & This Work \\
\vsini (km/s)                        & 2.4$\pm$0.2  & \citet{ivanyuk13} \\
Age (Gyrs)                              & 2$^{+3}_{-2}$   & This Work \\
Jitter - fit (m/s)                     &  0.90$^{+0.94}_{-0.33}$ & This Work\\

\hline
\end{tabular}

\end{table}

\subsection{Abundance Pattern}

One parameter where there is some difference between the solar value and that of HD41248 is the stellar metallicity.  The iron abundance ([Fe/H]) of 
this star is -0.43$\pm$0.10~dex, which is significantly lower than the solar value, defined as 0.0~dex.  In fact, this low abundance value could be one 
of the important parameters that defines a low-mass rocky dominated system from a gas giant dominated system, since the rising power law probability 
function that shows the most metal-rich Sun-like stars have a higher probability of hosting giant planets (\citealp{fischer05}; \citealp{sousa11}) seems to 
disappear or turnover for the rocky super-Earth population (\citealp{buchhave12}; \citealp{jenkins13a}).  Therefore, HD41248 can be classed as a 
metal-deficient and old solar analogue.

Table~\ref{tab:abund} lists various abundances for volatile elements in the atmosphere of HD41248.  We measured these values directly from the spectra 
using the method explained in \citet{pavlenko12} and more details of how we arrived at these values will be discussed in \citet{ivanyuk13}.  These 
abundances generally track the low [Fe/H] values, and in comparison to the Sun, all of the elements we have considered here are depleted.  This shows 
that the nascent disk from which planets could form was metal-deficient, are therefore depleted in the typical elements we expect are processed 
into planetesimals to form systems of planets through core accretion.  

\begin{table}
\center
\small
\caption{Chemical abundances for HD41248.}
\label{tab:abund}
\begin{tabular}{cccc}
\hline
\multicolumn{1}{c}{Element}& \multicolumn{1}{c}{[X/H]} & \multicolumn{1}{c}{[X/H]$_{\odot}$} & \multicolumn{1}{c}{[X/Fe]} \\ 
\multicolumn{1}{c}{}& \multicolumn{1}{c}{[dex]} & \multicolumn{1}{c}{[dex]} & \multicolumn{1}{c}{[dex]} \\ \hline
Si \sc i \rm  & -4.743$\pm$0.031 &  -0.34  &  0.09 \\
Si \sc ii \rm & -4.595$\pm$0.025 &  -0.20  &  0.24$^1$ \\
Ca \sc i \rm &  -5.972$\pm$0.017 &  -0.31 &  -1.14  \\
Ti \sc i \rm &  -7.419$\pm$0.017 &  -0.44 &  -2.59$^2$ \\
Ti \sc ii \rm & -7.326$\pm$0.032 &  -0.35 &  -2.49  \\
V  \sc i \rm  & -8.348$\pm$0.216 &  -0.31 &  -3.52$^1$ \\
Cr \sc i \rm  & -6.791$\pm$0.020 &  -0.35 &  -1.96  \\
Fe \sc i \rm  & -4.833$\pm$0.007 &  -0.42  &     -  \\
Fe \sc ii \rm & -4.845$\pm$0.018 &  -0.44  & -0.02 \\
Ni \sc i \rm &  -6.206$\pm$0.010 &  -0.39 &  -1.37 \\
\hline
\end{tabular}
\medskip

$^1$ - only a few absorption lines used in the analysis \\
$^2$ - significant non-LTE effects 

\end{table}

Finally, it is necessary to know the stellar mass of HD41248 so that we have a handle on the masses of any planets detected orbiting the 
star from the Doppler curve.  Given the \teff ~we measure of 5713$\pm$50~K, the absolute magnitude (M$_{V}$) of 5.22$\pm$0.08~mags, 
and the metallicity of -0.43$\pm$0.1~dex, we find a stellar mass of 0.92$\pm$0.05~\msol.  This is in agreement with other works who 
find similar values, both lower (0.81~\msol; \citealp{sousa11}) and higher (0.97; \citealp{casagrande11}).

\section{Doppler Analysis\label{rv}}

The radial velocity timeseries of 62 observations for HD41248 show evidence for at least two low-amplitude signals embedded in the 
data.  Both a periodogram analysis and our Bayesian fitting method detected these signals, however the Bayesian method can 
detect the second signal with a high degree of significance such that we can confirm the signal is robust.

\subsection{Periodogram Analysis}

\begin{figure}
\vspace{5.5cm}
\hspace{-4.0cm}
\includegraphics{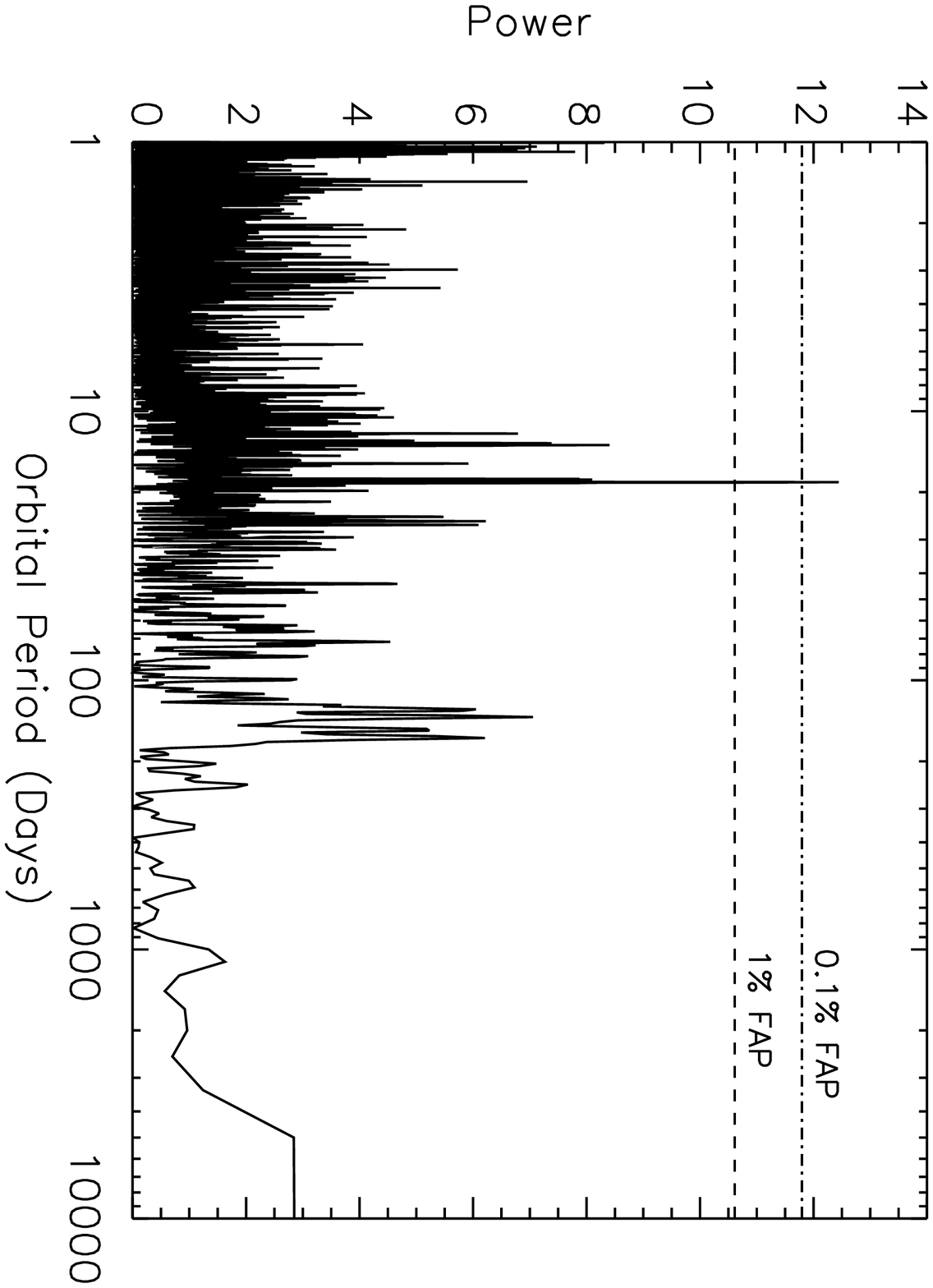}
\includegraphics{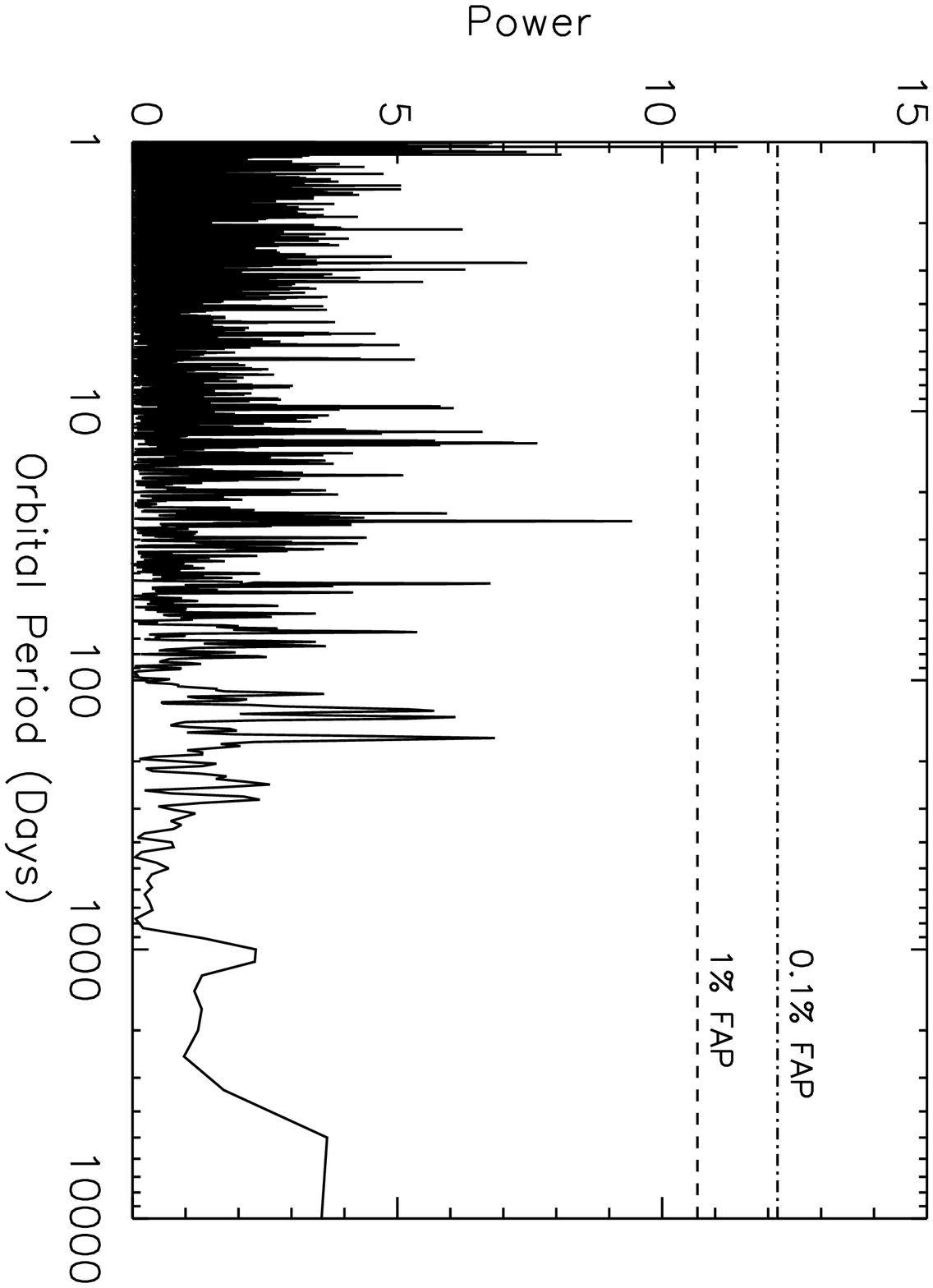}
\includegraphics{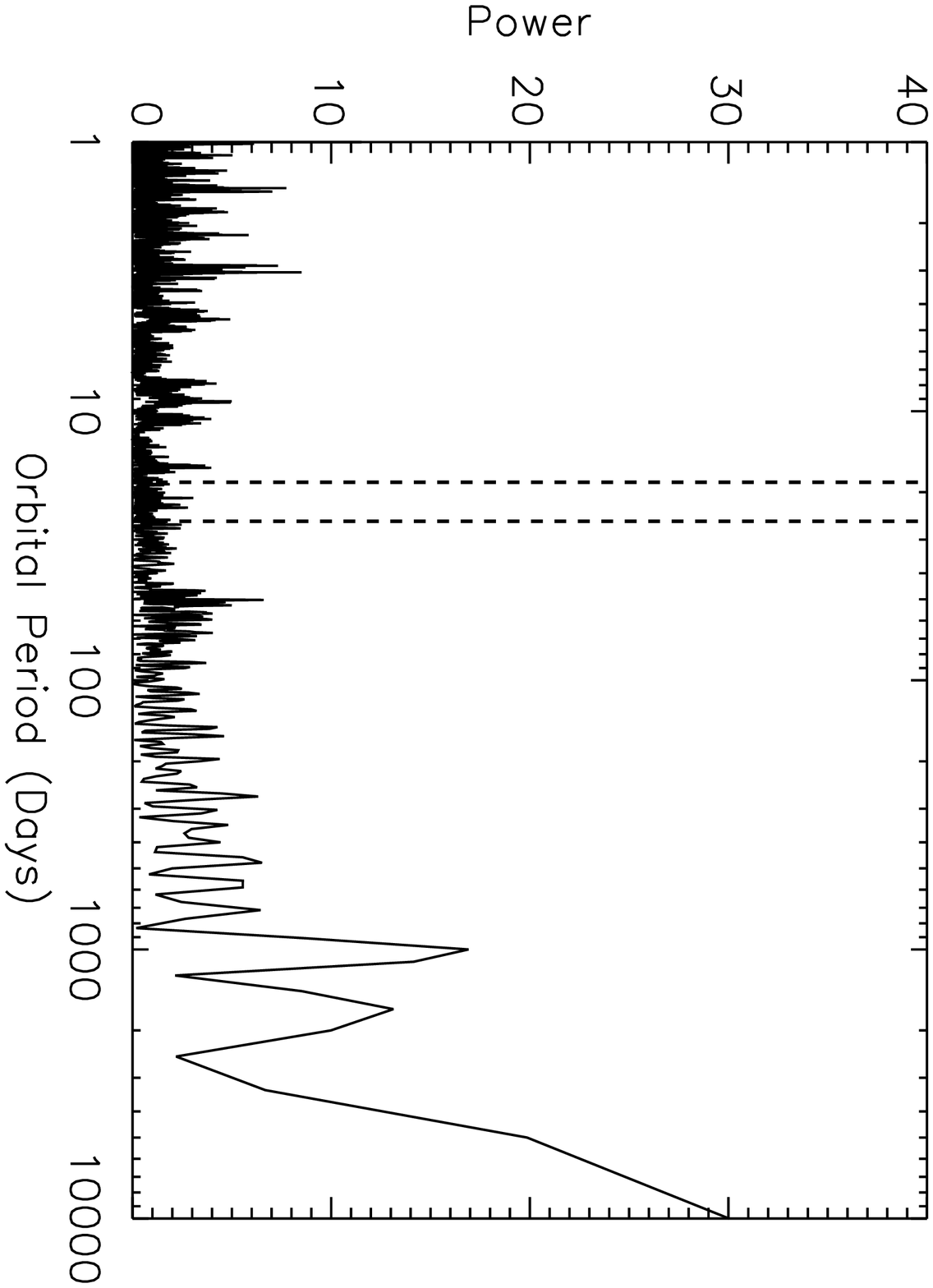}
\vspace{11cm}
\caption{Lomb-Scargle periodograms of the HD41248 radial velocity dataset is shown in the top plot, with the same analysis performed on the 
residuals to the 1-planet fit shown in the lower plot.  The dashed and dot-dashed lines mark the 1\% and 0.1\% FAP limits.  The lower 
plot shows the window function, along with two vertical dash lines that show the positions of the two signals.}
\label{period}
\end{figure}

A periodogram search for strong and stable frequencies in the radial velocity dataset for HD41248 reveals a significant peak around 
18~days.  The top panel of Fig.~\ref{period} shows these frequencies, with the horizontal dashed line marking the 1\% false-alarm probability 
(FAP) and the horizontal dot-dashed line marking the 0.1\% FAP, both were measured using a bootstrap analysis technique (see \citealp{anglada-escude12b}).  
It can be seen that the signal is stronger than both these boundaries.

The FAPs from the bootstrap analysis allow us to understand the significance of the frequency peaks we detect in the data.  Bootstrapping is 
appropriate in this case since we can generate test data sets from the original data set, without assuming any underlying distribution for 
the velocities, or more importantly, their uncertainties.  A lot of effort is being spent at the present time to understand how the combination 
of white and red noise affects the overall uncertainty we can assume for any radial velocity timeseries, particularly in HARPS data (\citealp{baluev13}; 
tuomi13b), however it is still an extremely difficult task to model such data with any high degree of accuracy.

We resampled the HD41248 data 10'000 times with replacement, using a Monte Carlo approach to scramble the timestamps of the velocities, but 
retaining the velocities and their associated uncertainties.  With each new sample, we recompute the LS periodogram and measure the strength of the 
strongest peak that we find.  The strength of this strongest peak is then compared to the strength of the original peak from the observed data set 
and the FAP relates to the number of times a peak stronger than this one is found.  In this way we can directly measure the probability from 
the data in an unparametric way.

The center panel in Fig.~\ref{period} shows the periodogram of the residuals to the best fit Keplerian of $\sim$18~days.  
There is a clear emerging signal around 25~days, indicating another physical process is causing a frequency peak at this period, 
however the significance is below the 1\% FAP level, meaning the archival data we have used is not yet abundant enough to confirm 
the nature of this peak. 

The lower plot in Fig.~\ref{period} shows the periodogram power for the widow function, with the strongest energies found at long periods, 
beyond the baseline of the data.  No sampling power is found around 18~days or 25~days, indicating the periods we detect in the radial velocities are 
not sampling features.  Also, no sampling power was found around 65~days, which is an alias that could arise since $\frac{1}{64.8} \approx \frac{1}{18.4} - \frac{1}{25.7}$, 
however we do find some activity power around 60-70~days that we discuss later but our Bayesian analysis indicates this is not the source of the secondary 
signal and it is a real Doppler shift, not an alias.  Such a tantilising system necessitates another methodology to confirm the significance of this signal and 
we turn to our Bayesian analysis method since this has been shown to efficiently detect significant signals with less velocity rich data 
than can be accomplished using a standard Lomb-Scargle periodogram analysis (\citealp{tuomi13a}).

\subsection{Bayesian Search}

Bayesian signal detection techniques have recently been introduced to spearhead the detection of radial velocity signatures of low-mass exoplanets around nearby stars \citep[e.g.][]{anglada-escude12c,tuomi12a,tuomi13b}, yet they have been shown to be rather immune to detections of false positives \citep[e.g.][]{tuomi11}. 
We analysed the HD41248 velocities using posterior sampling techniques and Bayesian model selection to find the best statistical descriptions, i.e. models, of the 
data, and to obtain estimates for the parameter probability densities of the corresponding models. Following \citet{tuomi12a}, we performed the samplings using 
the adaptive Metropolis algorithm \citep{haario01} and calculated the Bayesian evidences of each model using their truncated posterior mixture estimates. 
As we analysed the data in the Bayesian context, we defined the prior probability densities and model probabilities according to the choices of 
\citet{tuomi12b} and \citet{tuomi13b}.

When modeling the data, we used a common Gaussian white noise model as a reference model and attempted to improve this description by including correlations 
within it. This means that the measurement mean is described using a function $\mu(\theta, \xi, t)$, where $\theta$ is the parameter vector, $t$ represents time, and 
$\xi$ is a vector containing any other variables that might have an effect on the function. With this notation, we defined this model as
\begin{equation}\label{eq:mean_model}
\mu(\theta, \xi, t) = f_{k}(\theta_{p}, t) + \gamma + \sum_{i=1}^{p} c_{i} \xi_{i} ,
\end{equation}
where $f_{k}(t)$ is a function describing the superposition of $k$ Keplerian curves with orbital parameters $\theta_{p}$, $\gamma$ is a reference velocity w.r.t. the 
data mean, and parameters $c_{i}$ describe the linear dependence of the function on the variables $\xi_{i}$. We used the S-index, BIS, and FWHM as these variables 
to be able to take into account the correlations of the radial velocities with these activity-related indices.

At this point we can introduce the activity indicators that we use in our Bayesian model.  The BIS values and the FWHMs are taken 
directly from the HARPS-DRS output and details of their origin and usefulness can be found in \citet{queloz01} and \citet{santos10}.  The chromospheric 
$S$-indices have been measured following our own recipes, as mentioned in $\S$~\ref{sun}.  Using this procedure the uncertainties of these HARPS $S$-indices 
are found to be less than 1\%.

\begin{table}
\center
\caption{Model probabilities}
\label{tab:evidence}
\begin{tabular}{lllll}
\hline
\multicolumn{1}{l}{$k$}& \multicolumn{1}{l}{$\ln $ BE}& \multicolumn{1}{l}{$\ln $ BE} & \multicolumn{1}{l}{$P(k)$} & \multicolumn{1}{l}{$P(k)$} \\
\multicolumn{1}{l}{} & \multicolumn{1}{l}{(Ref)} & \multicolumn{1}{l}{(Cor)} & \multicolumn{1}{l}{(Ref)}  & \multicolumn{1}{l}{(Cor)}\\ \hline
\hline
0 & -158.1 & -154.9 & 5.3 $\times 10^{-13}$ & 6.0 $\times 10^{-12}$ \\
1 & -141.5 & -136.5 & 4.5 $\times 10^{-6}$ & 3.0 $\times 10^{-4}$ \\
2 & -128.5 & -127.7 & $\sim 1$ & $\sim 1$ \\
\hline
\end{tabular}
\tablecomments{~Log-Bayesian evidences and associated probabilities \\ 
for the Keplerian models from $k=0, ..., 2$ for the HD41248 \\
velocities taking into account the correlation terms with activity \\
indices (Cor) and without including these terms (Ref).}
\end{table}

\begin{figure*}
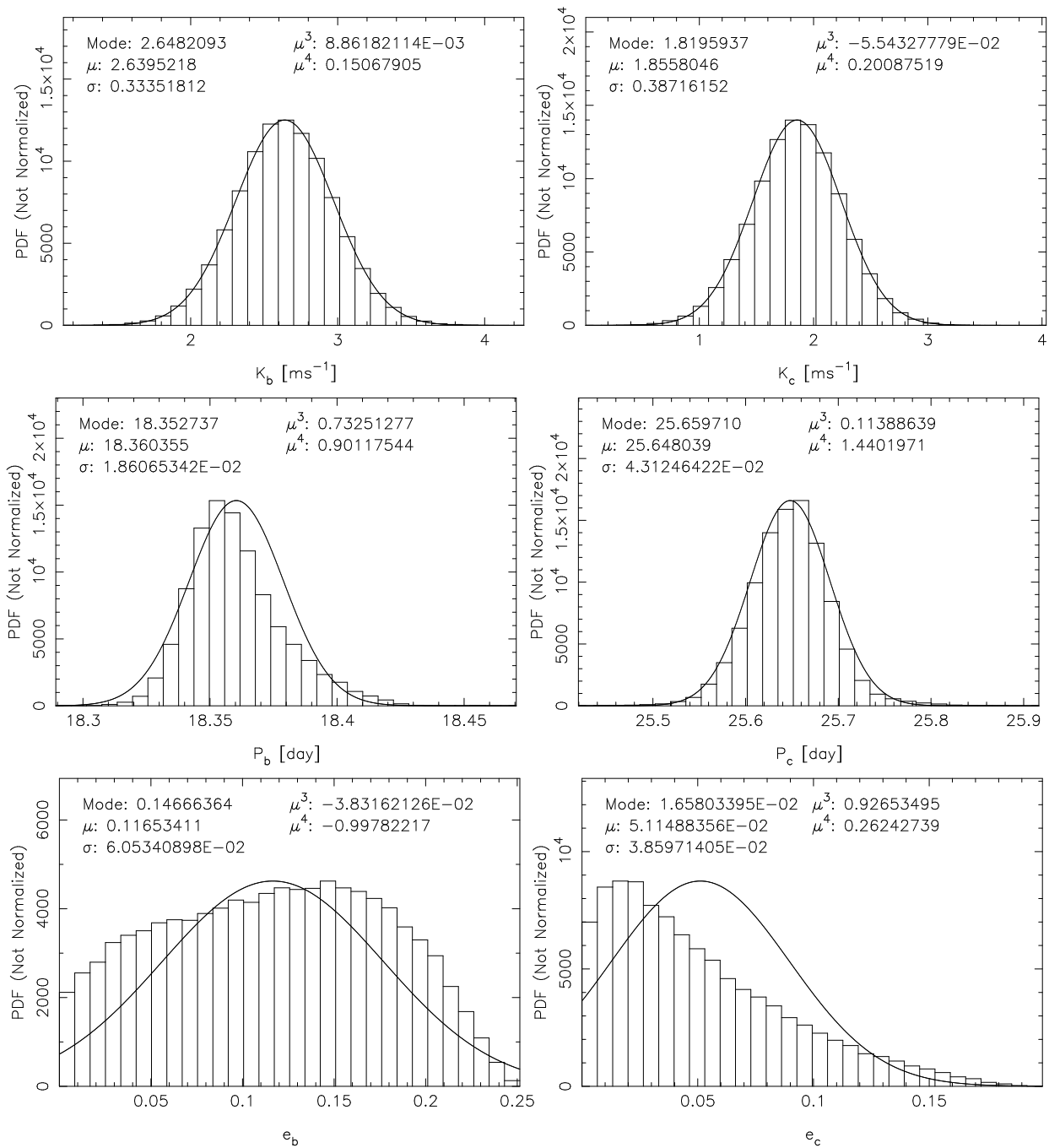

\vspace{0.0cm}
\hspace{-4.0cm}
\includegraphics{Fig8.ps}
\includegraphics{Fig9.ps}
\includegraphics{Fig10.ps}
\includegraphics{Fig11.ps}
\includegraphics{Fig12.ps}
\includegraphics{Fig13.ps}
\vspace{12.5cm}
\caption{The left column corresponds to the short period signal and the right column corresponds to the longer period 
signal.  The top two histograms show the estimated probability density function of periods from the Bayesian analysis, the 
middle histograms are the same for the semi-amplitude of the signals, and the lower two histograms are for the eccentricities 
of the signals.  Also overplotted by the solid curves are Gaussians with the same mean and sigma as the distributions.}
\label{bay}
\end{figure*}

Our results indicate that there are two significant periodicities in the HD41248 velocities at 18.4 and 25.6 days. We demonstrate the significance of the two signals 
by showing the log-Bayesian evidences of models with up to two Keplerian signals (Table~\ref{tab:evidence}). Accounting for 
the correlations between the velocities and the activity indices improves the model clearly for $k=0$ and $k=1$ but only marginally for $k=2$. Yet, the two-Keplerian 
model is clearly the best description of the data regardless of whether we account for these correlations or not and the second signal is detected according to the 
detection criteria of \citet{tuomi12b} because the two-Keplerian model is 3.3 $\times 10^{3}$ times more probable than the one-Keplerian model. The other two 
detection criteria, i.e. that the orbital periods and radial velocity amplitudes are well-constrained, are also satisfied. We demonstrate this by 
plotting the estimated posterior densities of the orbital periods, velocity amplitudes, and orbital eccentricities in Fig.~\ref{bay}. We have also tabulated our 
estimates for model parameters in Table~\ref{tab:system}.

\begin{table*}
\center
\caption{Keplerian solutions for HD41248.}
\label{tab:system}
\begin{tabular}{ccc}
\hline
\multicolumn{1}{c}{Parameter}& \multicolumn{1}{c}{HD41248~$b$} & \multicolumn{1}{c}{HD41248~$c$} \\ \hline
\hline
$P$ [days] & 18.357 [18.313, 18.423] & 25.648 [25.518, 25.768] \\
$e$ & 0.15 [0, 0.26] & 0.00 [0, 0.18] \\
$\omega$ [rad] & 3.2 [0, 2$\pi$] & 1.0 [0, 2$\pi$]$^{1}$ \\
$M_{0}$ [rad] & 0.2 [0, 2$\pi$] & 1.7 [0, 2$\pi$]$^{1}$ \\
$K$ [ms$^{-1}$] & 2.93 [1.65, 3.65] & 1.84 [0.67, 2.97] \\
$a$ [AU] & 0.137 [0.126, 0.154] & 0.172 [0.158, 0.192] \\
$m_{p} \sin i$ [M$_{\oplus}$] & 12.3 [6.9, 16.5] & 8.6 [3.6, 15.1] \\
\hline
$\gamma$ [ms$^{-1}$] &  -48 [-1.36, 0.18] & \nodata \\
$\sigma_{J}$ [ms$^{-1}$] & 0.90 [0.57, 1.84] & \nodata \\
$c_{1}$ [ms$^{-1}$ dex$^{-1}$] & 147 [-20, 296] & \nodata \\
$c_{2}$ & 0.000 [-0.232, 0.229] & \nodata \\
$c_{3}$ [10$^{-3}$] & 61 [-14, 156] & \nodata \\

\hline
\end{tabular}
\tablecomments{The maximum \emph{a posteriori} estimates of the solution for HD41248 
velocities and the 99\% Bayesian credibility intervals from posterior samplings.}
\end{table*}

Obtaining the samples from the posterior densities was simple in this case because the period-space contained only two significant maxima 
corresponding to the two signals we observed. Despite several attempts, we could not find a third signal and the samplings of the parameter 
space of a three-Keplerian model did not converge to a third periodicity. We also attempted to include a moving average (MA) component in 
the statistical model to improve its performance and to take into account intrinsic correlations in the measurement noise \citep{tuomi13b,tuomi13c}. 
However, the corresponding parameter describing the amount of autocorrelation in the data was found to be consistent with negligible estimates, 
and the corresponding MA component did not improve the statistical model. Furthermore, as can be seen in Table~\ref{tab:system}, all the parameters 
quantifying the linear dependence of the velocities on the different activity indices ($c_{1}$, $c_{2}$, and $c_{3}$) are consistent with zero (i.e. cannot 
be shown different from zero with a 99\% credibility), which explains why taking these correlations into account improved the statistical model little 
in terms of Bayesian evidences (Table~\ref{tab:evidence}). However, there are likely correlations between the velocities and S-index and FWHM, but 
BIS cannot be shown to be correlated with them at all.  

\begin{figure}
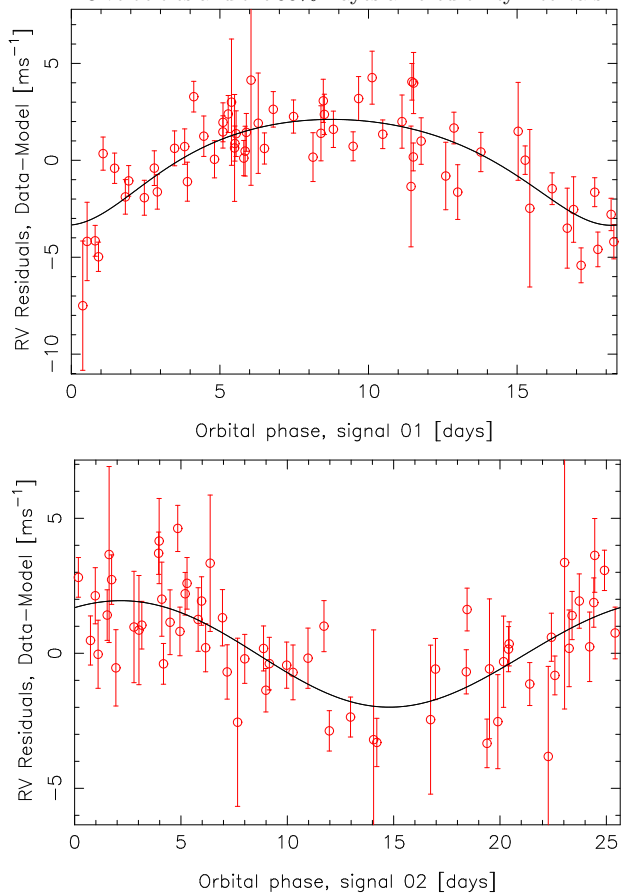

\vspace{5.0cm}
\hspace{-4.0cm}
\includegraphics{Fig14.ps}
\includegraphics{Fig15.ps}
\vspace{5cm}
\caption{Phased radial velocity curves for planets HD41248~$b$ and $c$ from the Bayesian detected periods.}
\label{orbits}
\end{figure}

Based on the Bayesian analyses of the HD41248 radial velocities, there are two significant periodicities in the data corresponding to two super-Earths 
or Neptune-mass planets on nearby close-in orbits (Fig.~\ref{orbits}). In fact, their orbits are so close to one another that the 99\% credibility 
intervals of $a_{b}$ and $a_{c}$ taking into account the uncertainty in the stellar mass are almost overlapping (Table~\ref{tab:system}) and the configuration 
cannot be immediately stated to be stable in the long term. However, the orbital stability is suggested by the fact that the ratio of the orbital periods is 
almost exactly 1.4 with a Bayesian 99\% interval of [1.384, 1.407], which suggests a possible 7:5 MMR.  The eccentricities of 
the system also show an interesting configuration with the inner planet having an eccentricity around 0.2 and the outer planet having an eccentricity close to 
circular.  In fact, we tested if the inner planet's eccentricity was closer to zero by changing the eccentricity prior model to 
favour a more circular orbit, but the eccentricity was found to be 0.22 in this case, indicating the inner planet does have some genuine measurable 
eccentricity given the data.  If we can rule out the source of these signals as originating from stellar magnetic activity, this could be a remarkable 
7:5 MMR system, and therefore we assess this possibility in $\S$~\ref{stability} by analysing the stability of the system.

\section{Activity Indicators\label{activity}}

Within our Bayesian model we have taken into account any linear correlations between the radial velocities and the activity indicators (BIS, FWHM, and $S_{\rm{HARPS}}$), 
as explained in the previous section.  However, not only does the presence of activity add additional noise to any individual radial velocity measurement, but it 
can also mimic the presence of a true Doppler induced frequency in the full timeseries of radial velocity data (e.g. \citealp{queloz01}).  Therefore, it is necessary to 
perform a test on whatever activity indicators one has at hand, to search for periodicities that could be associated with the periodicities found in the velocities.

\begin{figure}
\vspace{5.0cm}
\hspace{-4.0cm}
\includegraphics{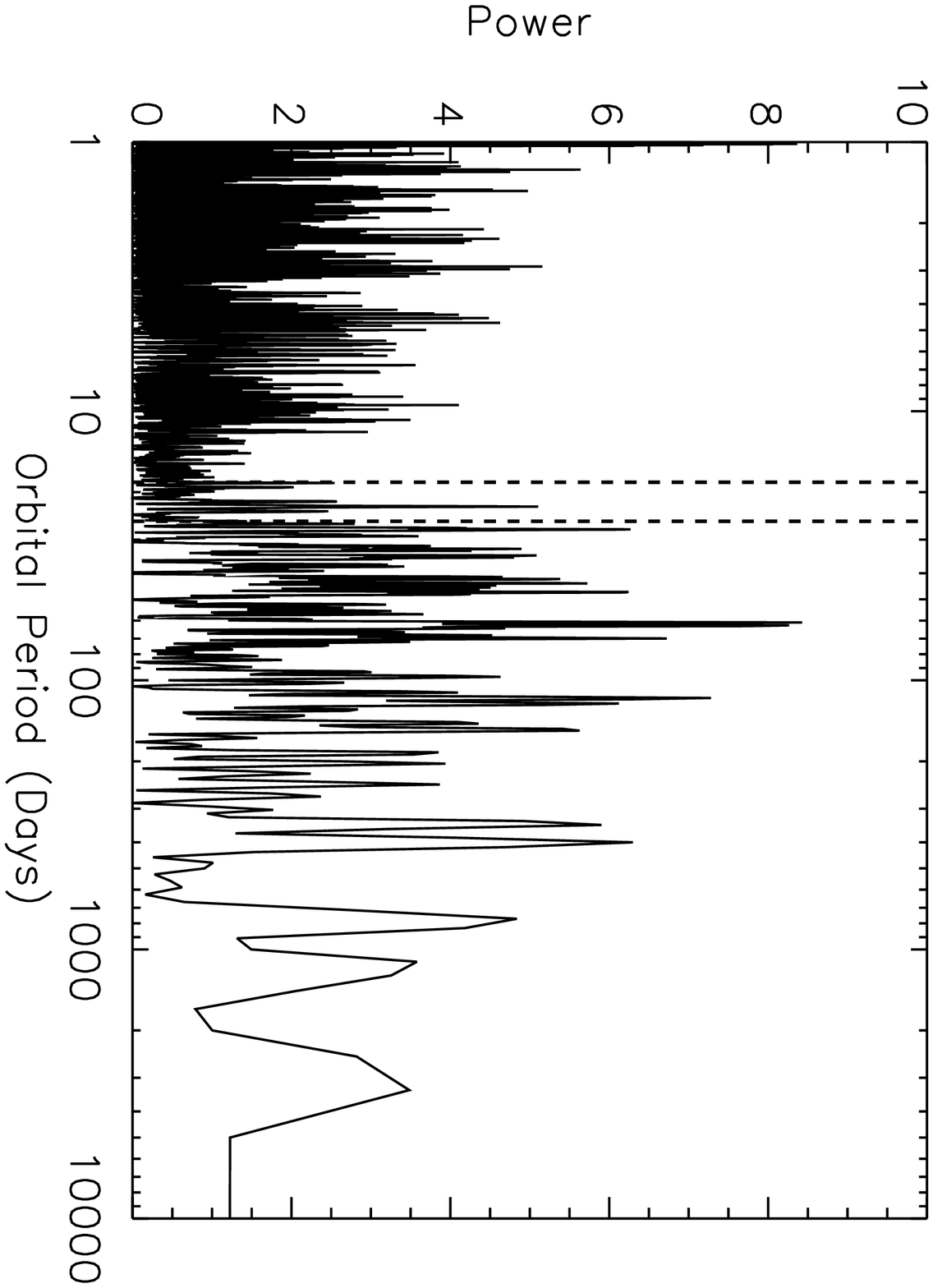}
\includegraphics{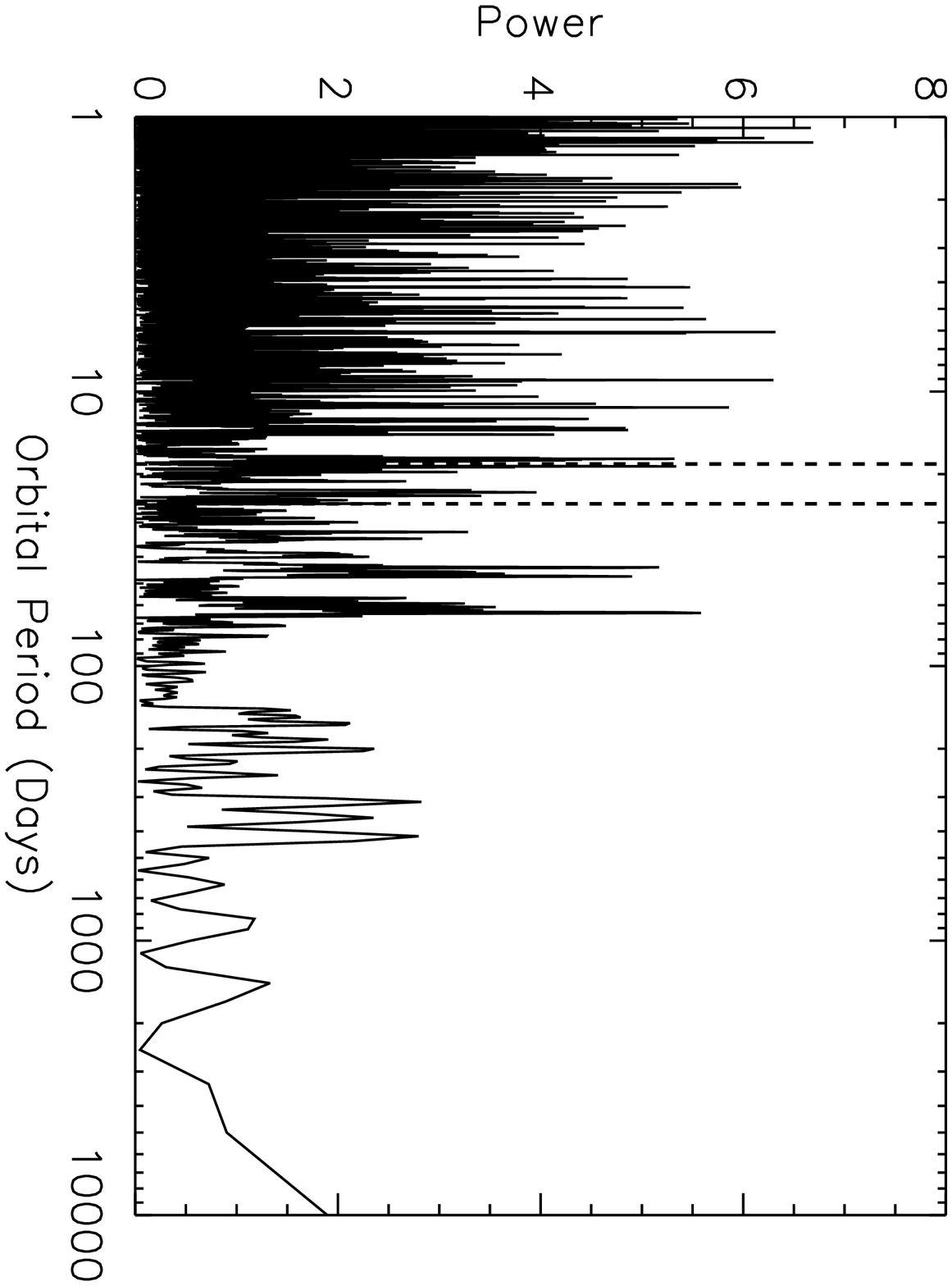}
\includegraphics{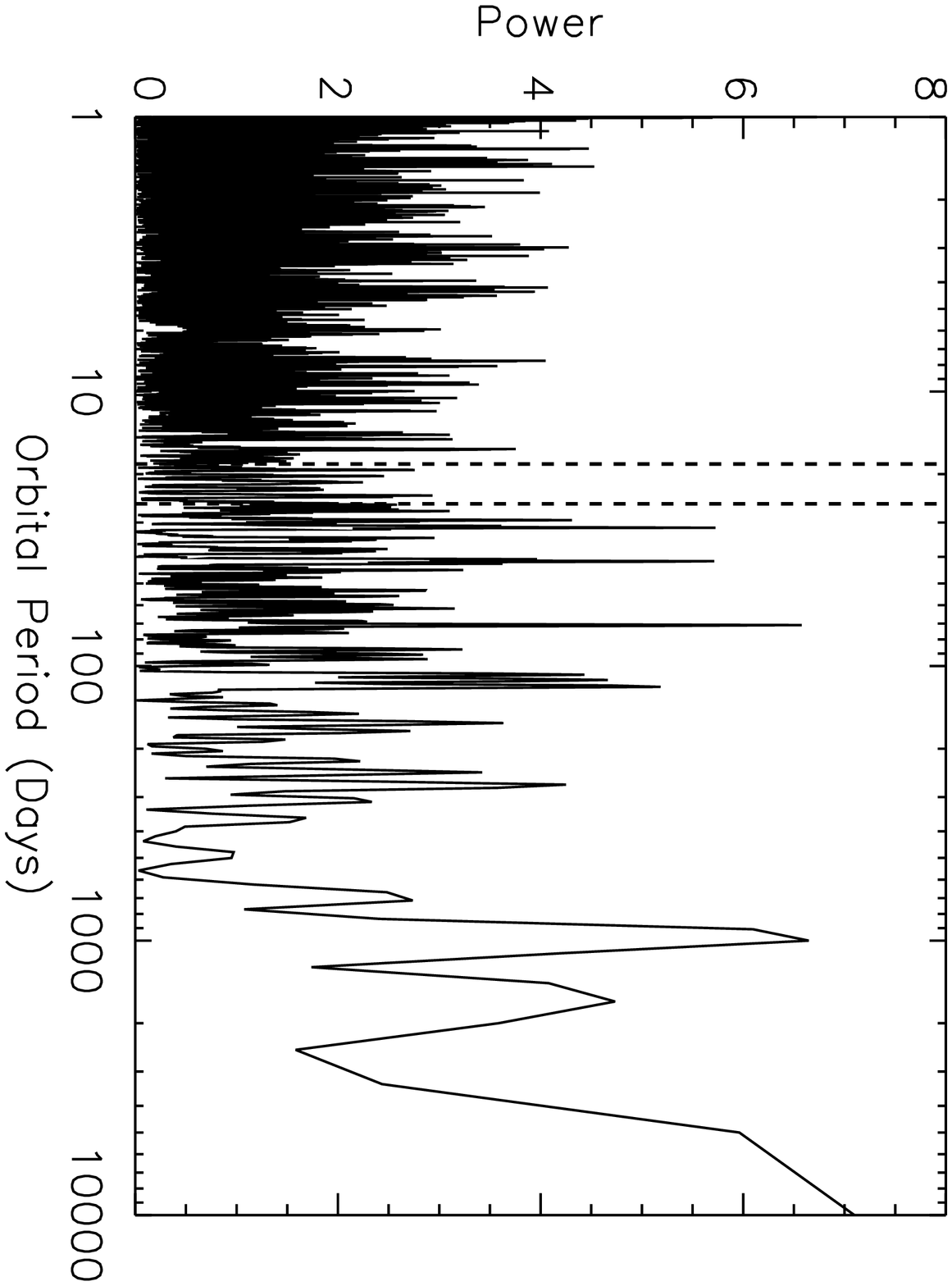}
\vspace{11cm}
\caption{Periodograms of the activity indices we consider.  From top to bottom we show the $S$-indices, BIS values, and the 
FWHM values.}
\label{act_peri}
\end{figure}

In Fig.~\ref{act_peri} we show the periodograms for each of these activity indices, going from the $S$-indices at the top, to the BIS values in the center, and finally the 
FWHMs at the bottom.  The $S$-indices show a strong periodogram peak at 61~days and another slightly weaker peak at 69~days which could be the rotational period 
of the star, or some other activity cycle.  There is no significantly strong frequency peak in the BIS periodogram, however in the FWHM periodogram there is a strong 
peak emerging at $\sim$71~days that could be related to the same frequency detected in the $S$-indices.  If so this may back-up the hypothesis that the rotational 
period of this star, or a harmonic thereof, is around 70~days, or at least there is an activity cycle present at that frequency.

The inner signal in the radial velocities at 18~days appears close to a peak in the BIS periodogram at around 18~days and this could throw some doubt on the nature 
of the inner signal, even though it is very unlikely for two signals to appear in a radial velocity dataset entrenched so close to such a MMR, down to the level of minutes.  
Also the peak in the BIS periodogram is not significant, with many more stronger peaks in the data, and since we included correlations between the BIS and the radial 
velocities, we can be confident this is not the source of the short period signal we detect.  We also note that any power in the activity indicators at a frequency that 
could give rise to an alias in a periodogram search for signals, will not be detected by our Bayesian analysis, since our selection must give rise to signals with amplitudes 
that are significantly different from zero.  This also protects our method from being confused by peaks that we see in the window function.  Therefore, we must conclude 
that the two signals we detect close to a 7:5 MMR are genuine Doppler signals, induced by the presence of two low-mass planets orbiting HD41248.

\section{System Stability\label{stability}}

The HD41248 system poses an interesting dynamical challenge because of the proximity of the 7:5 mean-motion resonance and the close
spacing of the two planets, both to each other, and to the central star. Here General Relativistic (GR) effects play a role in the dynamical
stability. Before moving on to perform a numerical search for stable configurations, we list several key properties of the system
that determine its stability.

\citet{chambers01} defined several measures that he used to quantify a system of planets, these are: the Angular Momentum Deficit (AMD; 
\citealp{laskar97}), the fraction of total mass in the most massive planet ($S_m$), a spacing parameter ($S_s$) that scales as the planet
to star mass ratio $\mu^{1/4}$ rather than the Hill relation of $\mu^{1/3}$ (\citealp{chambers96}), and a concentration
parameter ($S_c$) which measures how the mass is concentrated in an annulus. We supplement this with the measure stating the average
spacing in Hill radii ($S_H$). The values of these quantities for Venus and Earth, Jupiter and Saturn and HD41248 are listed in
Table~\ref{chamb}.

\begin{table}
 \begin{tabular}{c|ccccc}
  System & AMD & $S_m$ & $S_s$ & $S_c$ & $S_H$ \\ \hline \\
  VE & 8 $\times 10^{-4}$& 0.554 & 17.5&204.5& 26.2\\
  JS & 1.6 $\times 10^{-3}$ & 0.769 & 8.3 & 80.3 & 7.9 \\
  HD41248 & 3.4 $\times 10^{-3}$ & 0.560 & 6.9 & 411.0 & 8.5
 \end{tabular}
\caption{\citet{chambers01} stability quantities for the terrestrial planets, giant planets and HD41248.}
\label{chamb}
\end{table}

As one may see, the AMD of HD41248 is nearly twice as high as that of Jupiter and Saturn, and much higher than that of Venus and
Earth. Systems with higher AMD have the possibility to be more chaotic (\citealp{laskar97}) and have more opportunity to exchange it among
the planets. The relatively high AMD of HD41248 and the fact that one planet appears much more eccentric than the other suggests that
both eccentricities will fluctuate with a large amplitude, with one planet being at a minimum (c) when the other is at a maximum (b).

Given that both planets have a nearly equal mass it is no surprise that $S_d \sim 0.5$. More interesting is the spacing parameter. This
quantity is only $\sim$7 while for Venus and Earth it is about 18. Even for Jupiter and Saturn it is slightly larger. However, it is
interesting to see how $S_H$ scales as a function of the masses of the planets. We have $S_H =(a_c-a_b)/r_H$ where $r_H$ is the mutual
Hill radius i.e. $r_h=\frac{1}{2}(a_b+a_c)[(m_b+m_c)/3M_*]^{1/3}$. This can be solved for $a_b/a_c$ to give

\begin{equation}
\frac{a_b}{a_c} = \frac{2\Gamma-S_H}{2\Gamma+S_H},
\end{equation}

where $\Gamma = [3M_*/(m_b+m_c)]^{1/3}$. Setting $S_H=7$ and $m_b \sim m_c \sim 10$~$M_\oplus$ yields $a_b/a_c \sim 0.83$, but when
inserting the masses of Jupiter and Saturn one has $a_J/a_S \sim 0.58$. Thus, even though Jupiter and Saturn have a similar spacing
than HD41248 in terms of their mutual Hill radii, they are spaced father apart in relative semi-major axis ratio than the HD41248
system, implying that Jupiter and Saturn need a higher eccentricity to begin crossing their orbits than the planets of HD41248.
Indeed, for HD41248 the orbits begin to cross when both planets have an eccentricity near 0.1, so unless they are apsidally aligned
or protected by a resonance, they will become unstable. The nominal eccentricity of planet b is near 0.1, so the system is close to
instability and the stability may crucially depend on their phasing.

\begin{figure}
\resizebox{\hsize}{!}{\includegraphics[angle=-90]{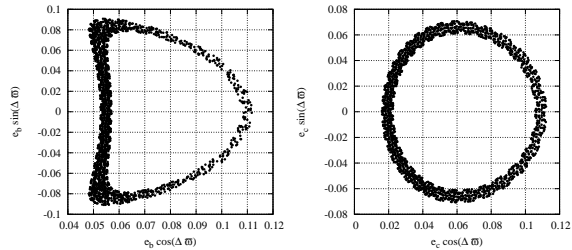}}
\caption{Secular evolution of the two planets of HD41248. The left panel shows the evolution of planet b in the $e-\Delta \varpi$
plane while the right panel shows the same for planet c. The planets are in apsidal alignment ($\Delta \varpi$ librates around 0).}
\label{hdevarpi}
\end{figure}

In Fig.~\ref{hdevarpi} we have plotted the secular evolution of both planets using the nominal orbital elements. The value of
$\omega_c$ is unconstrained so we set $\Delta \varpi = \omega_b-\omega_c$ i.e. the difference in the argument of periastron, to
20$^\circ$. The right panel shows the evolution of planet b in the eccentricity-$\Delta \varpi$ plane while the right panel shows the
same for planet c. The planets are in apsidal alignment i.e. $\Delta \varpi$ librates around 0 and both planets show substantial
excursions in their eccentricity amplitude. This alignment prevents many close encounters. We did not find the resonant angles to
librate for this simulation. The question is whether this motion persists for all initial conditions allowed by the observations. 
Numerical simulations should shed light on this issue.

\subsection{Numerical methods}

We proceeded to perform a large number ($\sim$ 40\,000) of short simulations of the HD41248 system, which were then analysed for
stability and or chaos.

We performed grid searches of both planets in the $a-\omega$ and $a-e$ planes. The stability dependence upon the mutual inclination
was not tested. At this stage the dependence on the masses was not tested either because of the large possible range depending
on the orientation of the system with respect to the observer. However, an estimate of the stability dependence on the masses can be
made and maximum masses can be computed as follows. 

For a planet on a fairly eccentric orbit, the chaotic boundary surrounding this planet is given by (\citealp{mustill12})

\begin{equation}
 \frac{\delta a}{a} \sim 1.8 (e\mu)^{1/5},
\end{equation}

This differs from the $\mu^{2/7}$ law of \citet{wisdom80} since it takes eccentricity into account. Taking both planets as
10~M$_{\oplus}$ and setting their eccentricities $\sim 0.05$ we have $\delta a/a ~\sim 0.12$. The planets are spaced 0.035~AU apart
and thus $\delta a/a < 1-a_b/a_c$, suggesting the system could be stable. Chaotic instability would set in when $\delta a/a \sim
1-a_b/a_c$, which can be solved and yields $e\mu \gtrsim 2 \times 10^{-5}$. Taking $e \sim 0.1$ implies that the masses need to exceed
60~M$_{\oplus}$, well outside the uncertainties.

We are dealing with two planets on planar orbits, so that we have four degrees of freedom: the semi-major axis or period ratio, the
eccentricities and $\Delta \varpi$. We performed a grid search for all of these parameters.

The first grid search was done in the semi-major axis-argument of periastron plane of planet b, keeping all other orbital elements at
their nominal values. The inclination and longitude of the ascending node were set to zero. For planet c Fig.~\ref{bay} shows that 
$\omega_c$ is almost arbitrary, so we set it to 0 for the sake of simplicity. We varied
$\omega_b$ evenly in steps of 3$^\circ$ between 0 and 360$^\circ$ and sampled $a_b$ assuming a Gaussian distribution with mean and
standard deviation given in Fig.~\ref{bay} using 99 steps. This resulted in 11880 simulations. The sampling
method employed here takes into account the density distribution of the semi-major axis but does not consider mutual correlations
between the elements.

The planets were integrated for 20\,500~yr using the SWIFT MVS integrator (\citealp{levison94}). We included the effects of GR 
by adding the potential term $V=-3(GM_*/c)^2a(1-e^2)/r^3$, where $c$ is the speed of light (\citealp{nobili86}). This
term reproduces the general relativistic effect of Mercury's perihelion advancement. For all simulations the time step was set to
2$\times$10$^{-4}$~yr and output was every 10~yr.

We analyse the stability of the system using frequency analysis (\citealp{laskar93}). The basic method is as follows: using a numerical
integration of the orbit over a time interval of length $T$, we compute a refined determination of the semi-major axes $a_1$, $a_2$
obtained over two consecutive time intervals of length $T_1 = T_2 =  T/2$. The stability index $D = max(\vert 1- a_2/a_1\vert)$
provides a measure of the chaotic diffusion of the trajectory. Low values close to zero correspond to a regular solution, while high
values are synonymous with strong chaotic motion (\citeauthor{laskar93}). The advantage of the frequency analysis method is that it does not
require long-term simulations and thus large regions of phase space can be tested with a reasonable amount of CPU power (\citealp{correia05}).

A similar methodology was employed for the other three sets of simulations. The eccentricities were sampled according to the best-fit
distributions and we set $\Delta \varpi=10^\circ$.

\subsection{Results}

In this section we present the results from the numerical simulations determining the stable regions. These results will be presented
in a series of figures.

\begin{figure}
\resizebox{\hsize}{!}{\includegraphics[angle=-90]{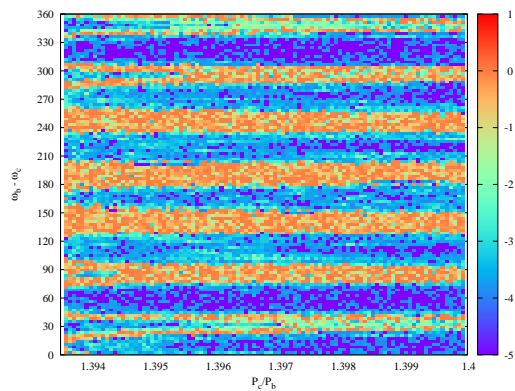}}
\caption{Stability map of the HD41248 system as a function of the period ratio (horizontal axis) and $\Delta \varpi$ (vertical axis).
The scale on the right indicates the value of the stability index $D$. Blue regions are stable, red is unstable. We varied $a_b$ and
$\omega_b$.}
\label{1ao}
\end{figure}

Fig.~\ref{1ao} displays the stability index of the system as a function of the period ratio $P_c/P_b$ and $\Delta \varpi$. For this
figure we varied the value of $a_b$ within its observed range. The axis labels indicate the quantity that was varied while the colour
coding shows $\log $D. Motion where $\log D \gtrsim 10^{-3}$ is considered chaotic and possible instability may occur (\citealp{correia05}).
 
An interesting feature are the seven horizontal regions of strong chaos and instability that appear independent of period ratio but
occur at regular intervals as $\Delta \varpi$ varies. We verified that in these regions the motion is unstable: the planets
encounter each other on short time scales. Varying $a_c$ and $\omega_c$ results in essentially the same outcome as that displayed in
Fig.~\ref{1ao}. The structure at low $\Delta \varpi$ and low period ratio marks the limits of the 7:5 resonance.

The reason for these regions of unstable motion have to do with the proximity of the 7:5 resonance. Without the resonance the system
would only have a single degree of freedom, $\Delta \varpi$, and thus be integrable (\citealp{michtchenko04}). The integrable
system may exhibit one of three types of motion: $\Delta \varpi$ librates around 0 (apsidal alignment), $\Delta \varpi$ circulates and
$\Delta \varpi$ librates around $180^\circ$ (apsidal anti-alignment; see \citet{morbidelli09} for a discussion on these three
types of motion). Without the presence of the resonance increasing the initial value of $\Delta \varpi$ from 0 to 360$^\circ$ results
in the system displaying all three types of motion in three distinct regions.

It should be noted that regions surrounding the period ratio of these planets are clustered with potential MMRs.  Nearby and stronger first-order MMRs are the 
3:2 and 4:3 commensurabilities, the similar strength 6:4 and 8:6 second-order MMRs around found around this dynamical region, and so are 
the weaker 10:7 and 11:8 third-order MMRs.  Therefore, we ran some tests to ensure the 7:5 MMR is the most likely dynamical configuration 
for this system, given the orbital parameters.  Taking the periods and uncertainties from Table~\ref{tab:system}, and assuming Gaussian 
distributions for each, we computed the distribution of the period ratio ($P_c$/$P_b$) and found a mean and $\sigma$ of 1.397 and 0.003, 
respectively.  The narrow range of periods given by the orbital elements constrains the possible MMR to be only the 7:5 ratio, since even the 
nearest 10:7 and 11:8 MMRs are found to be 8$\sigma$ and 6$\sigma$ away, respectively.

From the analysis of the motion of $\Delta \varpi$ we find that the system most likely has $\Delta \varpi$ librating
around 0 because this configuration yields all stable solutions. When $\Delta \varpi$ circulates or librates around $180^\circ$ the
system is chaotic, and often unstable.

However, the presence of the resonance changes the system's behaviour. The 7:5 is second order, and thus there are three resonant
arguments:

\begin{eqnarray}
 \sigma_b &=& 5\lambda_b - 7\lambda_c + 2\varpi_b, \nonumber \\
 \sigma_c &=& 5\lambda_b - 7\lambda_c + 2\varpi_c, \\
 \sigma_{bc} &=& 5\lambda_b - 7\lambda_c + \varpi_b + \varpi_c, \nonumber
\end{eqnarray}

where $\lambda = \varpi+M$ is the mean longitude and $\varpi = \Omega + \omega$ and $M$ is the mean anomaly. The system is
planar and thus $\Omega$ is undefined (we set it to zero). The proximity of the resonance causes the seven maxima and minima in $D$ as
a function $\Delta \varpi$ as it circulates or librates around 180$^\circ$. 

\begin{figure*}[b]
\resizebox{\hsize}{!}{\includegraphics[angle=-90]{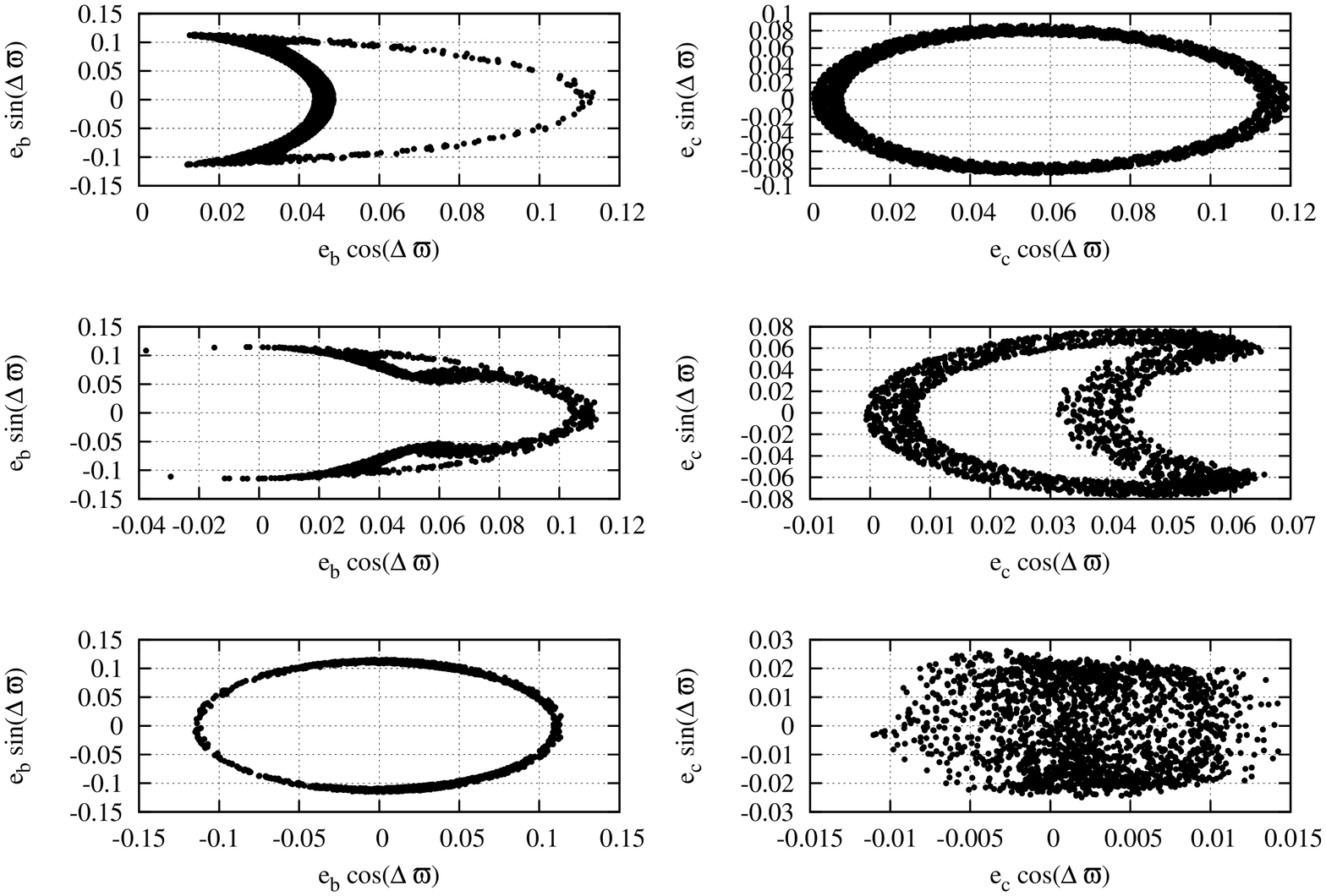}}
\caption{Evolution of $e_b-\Delta \varpi$ in the left column and $e_c-\Delta \varpi$ in the right column. The top panels show the
evolution before the instability, the middle panels are at the stability edge and the bottom panels show the evolution for an unstable
system.}
\label{trans}
\end{figure*}

We have decided to display this evolution in a series of figures. Starting from the bottom-right corner of Fig.~\ref{1ao} the system
becomes unstable when $\Delta \varpi \sim 30^\circ$ so a transition must occur. We displayed the evolution in the $e_b-\Delta \varpi$
(left column) and $e_c-\Delta \varpi$ planes (right column) for three simulations in Fig.~\ref{trans}. The top panels depict the
evolution while the motion is still regular, the middle panels show the motion at the edge of the instability strip, and the bottom
panels show the motion for an unstable configuration (although we only depicted the motion until the instability occurred).

In the top panels both planets have $\Delta \varpi$ librate around 0. Increasing the initial value of $\Delta \varpi$ from the top
panels to the middle panels one sees a fundamental change in the middle-left panel. There appear to be two regions of motion: one in
which $\Delta \varpi$ librates around 0 and one in which it circulates. The motion starts as libration around 0 with mean eccentricity
0.08 and small amplitude (small loop on the right) but the system encounters a saddle and can librate around 0 with large amplitude changes of the
eccentricity. The full motion is libration of $\Delta \varpi$ around 0 but it encompasses two separate regions. However,
every once in a while the separatrix is crossed and $\Delta \varpi$ also circulates. This suggests the system is on the edge of a
resonance. The motion for planet c has also changed: previously it showed a single loop but now it encompasses a moon-shaped region
that is typical for libration in a resonance (e.g. \citealp{murray99}). In the lowest panels $\Delta \varpi$ circulates because
the planets have crossed the separatrix. The apsidal no longer prevents them from encounters and the system is unstable.

We searched our solutions for libration of the angles $\sigma$ but we found no case where these angles librated. We checked this by
calculating the resonant angles for each simulation and tabulating the mean, minimum and maximum values. If libration occurs then the
minima and maxima should be different from 0 and 360 and the mean should be zero or 180. However, in each case we always found the
minima and maxima to be 0 and 360, and the mean to be close to 180, suggesting that the angles circulate. We have inspected a large
sample by eye and conclude that we did not witness any libration occurring. While this procedure should break down for unstable cases,
configurations which are stable for the entire simulation duration should have their minima and maxima confined to a small range. We
believe this could be caused by the influence of GR, which primarily affects the pericentre precession of close-in planets. In
turn, this can help break or support a resonant configuration and also affect the stability of multi-planet systems (e.g. \citealp{veras10}).

\begin{figure}
\resizebox{\hsize}{!}{\includegraphics[angle=-90]{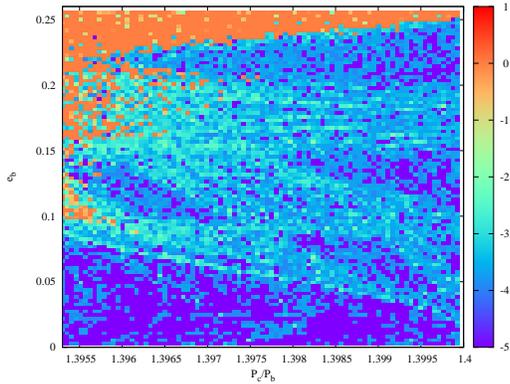}}
\caption{Stability map of the HD41248 system as a function of the period ratio (horizontal axis) and $e_b$ (vertical axis).
The scale on the right indicates the value of the stability index $D$. Blue regions are stable, red are unstable. We varied $a_b$ and
$e_b$.}
\label{1ae}
\end{figure}

\begin{figure}
\resizebox{\hsize}{!}{\includegraphics[angle=-90]{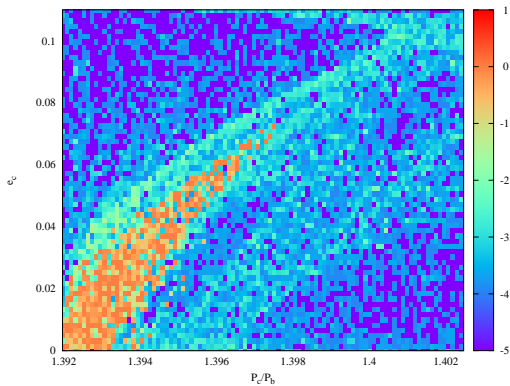}}
\caption{Same as Fig.~\ref{2ae} but now we varied $a_c$ and $e_c$.}
\label{2ae}
\end{figure}

The stability dependence of the system on the eccentricities of both planets is displayed in Figs.~\ref{1ae} and~\ref{2ae}. For these
simulations we set the initial value of $\Delta \varpi=10^\circ$ so that we would not witness any instability related to this
quantity. We varied $a_b$ and $e_b$ for Fig.~\ref{1ae} and $a_c$ and $e_c$ for Fig.~\ref{2ae}. As one may see, the system is mostly
stable as a function of $P_c/P_b$ and $e_b$ unless $e_b \gtrsim 0.2$, apart from some regions at low period ratio. For such high values
of $e_b$ the apsidal alignment does not protect the planets from encounters because the orbit of planet b already crosses that of
planet c, regardless of the eccentricity of planet c. There is some fine structure visible in the figure of which the line going from
(1.3955,0.08) to (1.4,0.03) is the most interesting. Below this line there is a large region of stable motion and here $\Delta \varpi$
circulates. However the eccentricities are small enough that the planets do not cross their orbits and they are protected by the 7:5
resonance. Above this region the stability is guaranteed by the apsidal alignment (apart from at high eccentricity).

In Fig.~\ref{2ae} we see a large unstable region at low eccentricity and low period ratio. Here the planets are at the edge of the
resonance which results in chaotic motion and large excursions in eccentricity resulting in encounters. The stable region on the right
has the motion being dominated by small eccentricity and by the resonance so there are few encounters, while at high eccentricity and
low period ratio the planets are not in resonance but the apsidal alignment prevents close encounters.

In summary, a system of two super-Earth planets is most likely in the HD41248 data when one considers the nominal orbits, provided that the
apses are nearly aligned. Given that $\omega_c$ appears unconstrained by observations, this configuration is a viable outcome.
The apsidal alignment provides protection against close encounters at high and low eccentricities, as long as the eccentricities are
of comparable value. If the eccentricity of one planet exceeds 0.2 while the other is nearly zero, the system becomes unstable.
However, such extreme configurations can be ruled out by the observational uncertainties.

\section{Discussion\label{discussion}}

The population of low-mass exoplanets tends to differ significantly from that of more massive planets.  Observational evidence for a 
lack of low-mass planets orbiting the most metal-rich stars (\citealp{jenkins13a}) points to fundamentally different evolutionary 
properties between low and high mass planets.  As discussed in the introduction, there also appears to be a higher fraction of dynamically 
packed low-mass planetary systems with planets close to MMR's.

Here we report the discovery of a possible 7:5 MMR planetary system, consisting of two super-Earths orbiting a metal-poor star.  Alone this 
discovery adds to the number of metal-poor stars with low-mass planets and lends more weight to the hypothesis presented in \citeauthor{jenkins13a}  
It could also be the first confirmed super-Earth system of planets in such a second-order resonant configuration since libration of the longitude 
of periastrons yield a highly stabilized system, and our data set and simulations leave the possibility for libration of the resonant angles, and hence 
resonance.  Also the Doppler data period ratio differs from the true 7:5 period ratio by less than 1 part in 300, a strong argument in favour of the MMR.

Around 1/3rd of multi-planet systems discovered by radial velocities are found to be close to a MMR (\citealp{lissauer11}) but the fraction actually 
in the resonance is significantly lower.  However, \citeauthor{lissauer11} show that the number of known systems in a MMR is significantly more 
than that expected from a randomly drawn population.  Predictions from convergent migration show that around 1\% of resonant systems should last 
for around a disk lifetime (\citealp{adams08}) whereas planet-planet scattering mechanisms yield MMRs around 5-10\% of the time (\citealp{raymond08}).  
If we take the Kepler resonant numbers from \citeauthor{lissauer11} at face value, since the number of false positives from multi-transiting systems is 
expected to be significantly lower than that from single transiting events (\citealp{latham11}), then we might expect that the HD41248 planetary system was 
formed through the planet-planet scattering mechanism.  However, it may not be so straight-forward.

\citet{terquem07} studied the migration of cores through a proto-planetary disk that includes mutual interactions between the cores and found that 
the presence of MMRs is common.  They found that simulations that began with a larger outer planetary distribution radius, or that started with significantly lower core masses, 
would tend to produce the 7:5 MMR.  This indicates that such a second-order MMR that forms in this way requires longer timescales, since reducing the core 
masses or increasing the planetary core distance from the star gives rise to a longer evolutionary time, and thus if the HD41248 planets were formed in 
this way we might expect they started with low initial core masses, possibly by forming late in the disk evolution, or they started life much further out in 
the disk than their current semi-major axes, or most likely a combination of both low core masses and long convergent migration.  It is interesting to note 
that even though planet-planet scattering simulations produce more MMRs, the results from \citet{raymond08} never resulted in a 7:5 MMR commensurability.

Finally, we should note the mass ratio between these planets.  HD41248~$b$ is found to be 12.3~\me and planet $c$ is 8.6~\me, giving rise to a mass ratio of 0.7.  The 
simulations of \citet{terquem07} ended with much lower mass ratios for the planets that were found in the 7:5 configuration.  They found mass ratios of 0.1 and 0.2, 
however the more massive planet was the one closer to the star, similar to the system we present here.  We do note that finding MMRs that have much lower 
mass ratios is difficult due to the inherent difficulties of first discovering very low-amplitude signals in radial velocity datasets, and also the difficulty of finding 
low-amplitude multi-planet signals when the super-position of the Keplerians has a strong dominating signal and a much weaker and longer period signal.  With 
this in mind, we expect more of these systems to emerge from the radial velocity data with the addition of more Doppler data and better analysis techniques.

\acknowledgments

We would like to thank the anonymous referee and the editor, Eric Feigelson, for their helpful comments that made this a better manuscript.  
We would also like to thank Guillem Anglada-Escud{\'e} for providing us with the HARPS-TERRA code.  
 JSJ also acknowledges funding by Fondecyt through grant 3110004 and partial support from CATA (PB06, Conicyt), the 
GEMINI-CONICYT FUND and from the Comit\'e Mixto ESO-GOBIERNO DE CHILE.   This research has made use of the SIMBAD database and the VizieR catalogue 
access tool, operated at CDS, Strasbourg, France.

\bibliographystyle{aa}
\bibliography{refs}

\end{document}